\documentclass[aps,prl,reprint,amsmath,amssymb,superscriptaddress,showpacs]{revtex4-1}
\usepackage{bm}
\usepackage{graphicx}
\usepackage{color}
\usepackage{dcolumn}

\begin{document}

\title{Magnon-phonon coupling and two-magnon continuum in two-dimensional triangular antiferromagnet CuCrO$_2$}

\author{Kisoo Park}
\affiliation{Center for Correlated Electron Systems, Institute for Basic Science (IBS), Seoul 08826, Republic of Korea}
\affiliation{Department of Physics and Astronomy, Seoul National University, Seoul 08826, Republic of Korea}

\author{Joosung Oh}
\affiliation{Center for Correlated Electron Systems, Institute for Basic Science (IBS), Seoul 08826, Republic of Korea}
\affiliation{Department of Physics and Astronomy, Seoul National University, Seoul 08826, Republic of Korea}

\author{Jonathan C. Leiner}
\affiliation{Center for Correlated Electron Systems, Institute for Basic Science (IBS), Seoul 08826, Republic of Korea}
\affiliation{Department of Physics and Astronomy, Seoul National University, Seoul 08826, Republic of Korea}

\author{Jaehong Jeong}
\affiliation{Center for Correlated Electron Systems, Institute for Basic Science (IBS), Seoul 08826, Republic of Korea}
\affiliation{Department of Physics and Astronomy, Seoul National University, Seoul 08826, Republic of Korea}
\affiliation{Laboratoire Leon Brillouin, CEA-Saclay, 91191 Gif-sur-Yvette Cedex, France}

\author{Kirrily C. Rule}
\affiliation{Australian Nuclear Science and Technology Organisation, Lucas Heights, 2234, New South Wales, Australia}

\author{Manh Duc Le}
\affiliation{Center for Correlated Electron Systems, Institute for Basic Science (IBS), Seoul 08826, Republic of Korea}
\affiliation{Department of Physics and Astronomy, Seoul National University, Seoul 08826, Republic of Korea}
\affiliation{ISIS Facility, STFC, Rutherford Appleton Laboratory, Didcot, Oxfordshire OX11-0QX, United Kingdom}

\author{Je-Geun Park}
\email{jgpark10@snu.ac.kr}
\affiliation{Center for Correlated Electron Systems, Institute for Basic Science (IBS), Seoul 08826, Republic of Korea}
\affiliation{Department of Physics and Astronomy, Seoul National University, Seoul 08826, Republic of Korea}

\date{\today}

\begin{abstract}
CuCrO$_2$ is a manifestation of a two-dimensional triangular antiferromagnet which exhibits an incommensurate noncollinear magnetic structure similar to a classical 120$^{\circ}$ ordering. Using the inelastic neutron scattering technique, direct evidence of a magnon-phonon coupling in CuCrO$_2$ is revealed via the mixed magnon-phonon character of the excitation mode at 12.5 meV as well as a minimum at the zone boundary. A simple model Hamiltonian that incorporates an exchange-striction type magnon-phonon coupling reproduces the observed features accurately. Also, continuum excitations originating from the interaction between quasiparticles are observed with strong intensity at the zone boundary. These features of the magnetic excitations are key to an understanding of the emergent excitations in noncollinear antiferromagnetic compounds.
\end{abstract}
\maketitle

\section{I. Introduction}
Geometrically frustrated antiferromagnets have been an extremely fruitful subject in condensed matter physics due to the various non-trivial macroscopic degeneracies originating from competing magnetic interactions and low dimensionality. A typical example of this state of matter is the widely studied low-temperature magnetism of two-dimensional triangular lattice antiferromagnets (2D-TLA). One such system is CuCrO$_2$, where layers of edge-shared CrO$_6$ octahedra, with the magnetic Cr$^{3+}$ ions forming a triangular lattice, are separated by layers of nonmagnetic Cu$^+$ ions \cite{seki2008spin} such that there is little magnetic coupling between the Cr layers. The octahedral environment of the Cr$^{3+}$ ($3d^3$) ions leads to a quenched orbital moment and total spin of 3/2 since the $t_{2g}$ orbitals are each half filled. A ‘proper helix’ magnetic ground state with a propagation vector \textbf{Q} = (0.329, 0.329, 0) occurs below T$_N \approx 24 $ K, which is close to a classical 120$^{\circ}$ ordering \cite{poienar2009Ndiff}.

Despite its well-known magnetic properties and earlier reports of inelastic neutron scattering (INS) experiments, the complete spin Hamiltonian of CuCrO$_2$ is still unresolved. One reason is due to some contradictory observations in the various INS studies. For instance, evidence for a flat 5 meV magnon-mode at the zone center is given in  \cite{Poienar2010INS, FrontzekINS2011}. However, this mode is clearly absent in the more recent work of Kajimoto et. al, \cite{Kajimoto2015INSdiffusion} as well as in this work. This discrepancy may be due to the difference in energy resolution between the experiments. Taken together, these observations negate the need to introduce an overly large single-ion anisotropy (SIA) in order to explain the unusual downward shift in the 5 meV mode at the zone boundary as was previously done. Instead, we show that this and other features of the data can be readily explained by taking into account the effects of a magnon-phonon coupling.

Making use of new INS measurements on CuCrO$_2$ single crystals, we have determined accurately the two key places where there are extra features in the observed spectra when compared to a pure Heisenberg model: (1) an additional intensity below T$_N$ in the phonon mode at 12.5 meV, (2) a minimum of the magnon dispersion at the zone boundary. In order to explain these two discrepancies, we present a refined model using the exchange-striction mechanism to account for a magnon-phonon coupling arising from the non-collinear magnetic structure in this compound. Our data also reveal a  continuum excitation in the magnon-phonon spectrum above the single magnon modes. We will discuss this continuum excitation and show that it orginates from the magnon-magnon interaction as well as the possible mixing of phonon and magnon modes.

\section{II. Experimental Details}
CuCrO$_2$ is a member of the delafossite (CuFeO$_2$) family with space group R$\overline{3}$m, a $\approx$ 2.97 and c $\approx$ 17.10 Å \cite{poienar2009Ndiff}. For this experiment, sizable single crystals of CuCrO$_2$ were grown by a flux decomposition method using a mixture of K$_2$Cr$_2$O$_7$ and CuO, following ref. \cite{Ye1994Growth}. Four single crystals of total mass $\sim$ 0.8 g were co-aligned on an aluminum sample holder by super glue with mosaicity of less than 1$^{\circ}$. These samples were mounted in the scattering plane (1, 1, 0) and (1, 0, 0) such that the wave vector of the form $\textbf{q}$ = (H, K, 0) reciprocal lattice unit is accessible in the horizontal plane.

Our INS experiment was carried out on the TAIPAN thermal triple-axis spectrometer \cite{danilkin2009taipan} at the Australian Nuclear Science and Technology Organisation (ANSTO), Australia. Energy scans with constant $\textbf{q}$ along the [H, H, 0] direction were performed by varying the incident neutron energy with a fixed final neutron energy of 14.87 meV giving an energy resolution of less than 1 meV. The pyrolytic graphite (002) monochromator and analyzer were used with collimation open $-$ 40$^{\prime}$ $-$ 40$^{\prime}$ $-$ open to optimize the beam intensity and the resolution in momentum and energy simultaneously. To cut out higher order scattering signals, a pyrolytic graphite filter was mounted between the sample and analyzer. The samples were loaded inside the bottom-loading cryofurnace CF-1, and measurements were made at its base temperature of 5 K to get a clear picture of the magnetic excitations. Additional scans at 50 and 300 K were carried out to investigate the temperature dependence of the phonon mode.
\begin{figure}
\includegraphics[width=\columnwidth,clip]{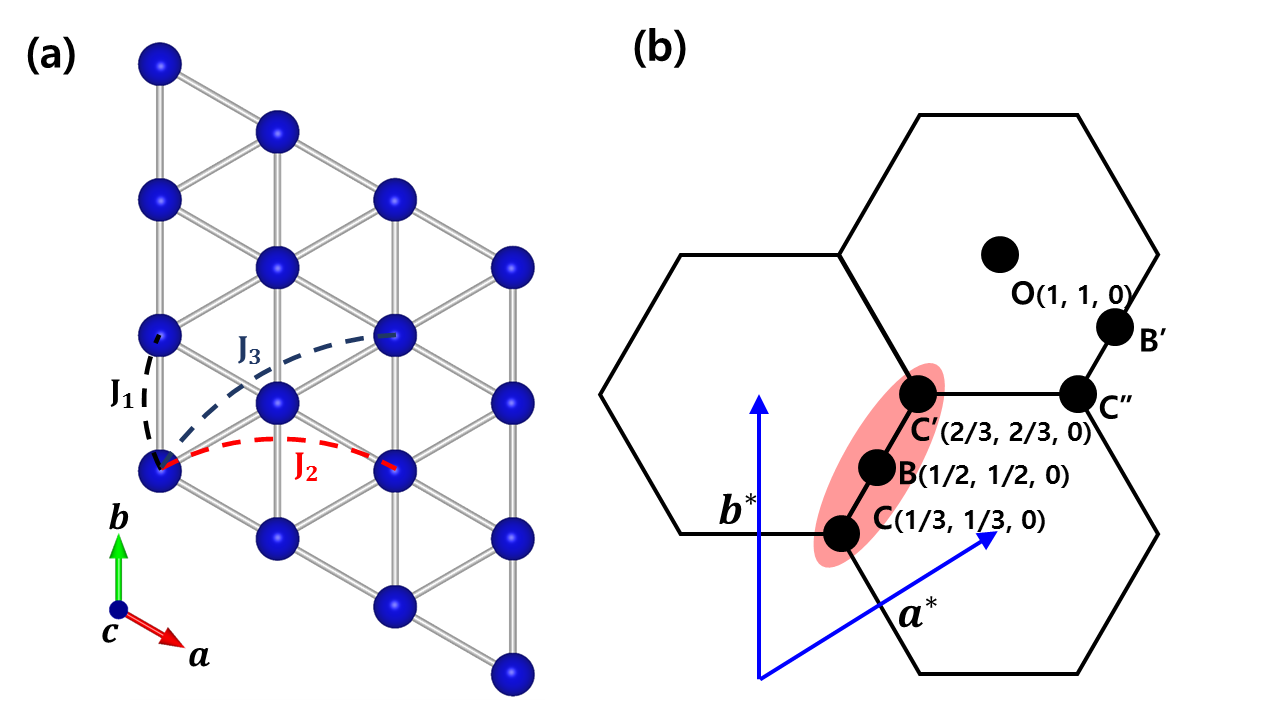}
\caption{ \label{fig1} (Color online) (a) Diagram of the Cr atoms in a 2D triangular plane with the exchange interactions between Cr atoms. (b) Diagram showing our E-scan regions in the 2D reciprocal lattice of CuCrO$_2$.}
\end{figure}

\section{III. Theoretical modeling of Spin Hamiltonian}
\subsection{A. Pure Heisenberg Hamiltonian}
\begin{figure*}
\centering
\includegraphics[width=\textwidth,clip]{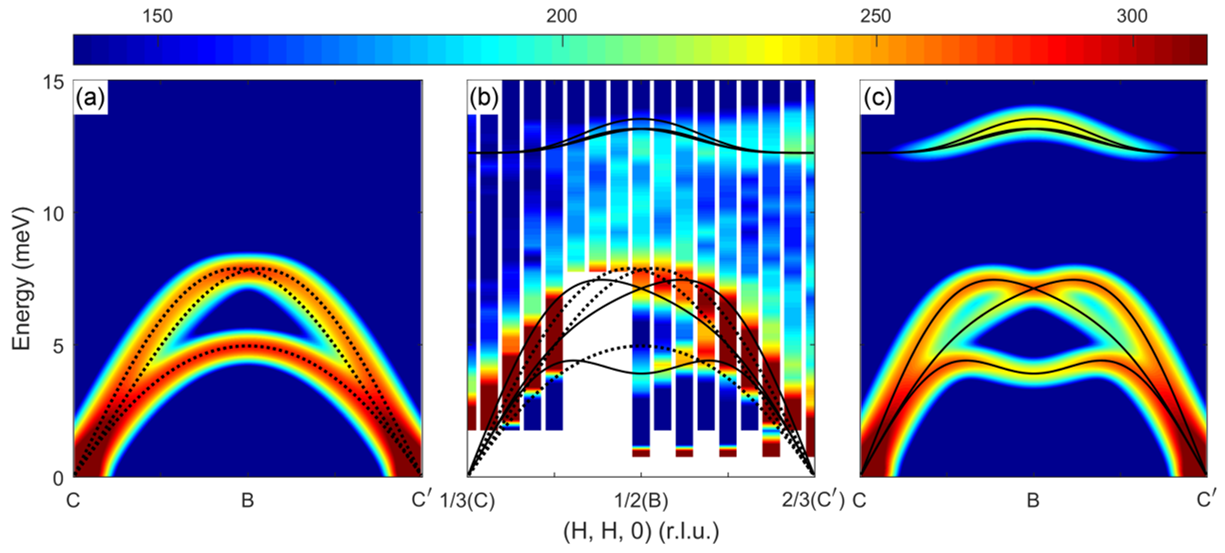}
\caption{\label{fig2}(Color online) Theoretically calculated single-magnon dispersion curve and dynamical structure factor S$(\textbf{q},\omega)$ using (a) the minimal Heisenberg Hamiltonian of Eq. $(\ref{hamiltonian})$ (dotted line) and (c) the full Hamiltonian Eq. $(\ref{hamiltonian2})$ (solid line) with magnon-phonon coupling after being convoluted using the Gaussian functions with fixed width of 0.59 meV. (b) Contour plots of the INS intensity of CuCrO$_{2}$ at T = 5 K along the [H, H, 0] direction in the reciprocal space. Solid lines at 12.5 meV denote the mixed magnon-phonon character of the excitation mode calculated using Eq. $(\ref{hamiltonian2})$.}
\end{figure*}

The map of the inelastic neutron scattering intensity $S(\textbf{q},\omega)$ along [H, H, 0] observed on TAIPAN is shown in Fig. \ref{fig2}(b). The overall shape of the single-magnon spectrum below 8 meV is consistent with previous studies \cite{Poienar2010INS,frontzek2011NDiff,Kajimoto2015INSdiffusion}. However, as stated above, we did not observe the flat mode at 5 meV near the zone center (C and C$^{\prime}$ point) reported earlier. Rather in our data, the weak scattering at the zone center appears to originate from the long tail of the elastic signal. These data are also in good agreement with the data presented in \cite{Kajimoto2015INSdiffusion}. Furthermore, the large SIA required to produce this flat mode is unlikely since only a small easy-plane anisotropy value is required to stabilise the observed magnetic structure, and this small anisotropy has little effect on the calculated magnon spectrum.

Even though there is a small component of an incommensurate magnetic structure present, the magnetic ground state of CuCrO$_2$ can be well approximated by a classical 120$^{\circ}$ ordered state which is typical in a Heisenberg TLA. Therefore, to maintain reasonable simplicity in our modeling, we neglect all of the interactions that are associated with the incommensurate magnetic structure of CuCrO$_2$, i.e. the interlayer coupling or the anisotropic exchange interaction. This is done in order to focus on fitting the overall shape of the magnon dispersion in $(\textbf{q}, \omega)$ space. Thus, as a starting point for analyzing the data, the following minimal Heisenberg Hamiltonian is utilized:
\begin{equation}
{\cal H}_{Heis} =  J_1\sum_{NN}{\textbf{S}_{i} \cdot \textbf{S}_{j}}+J_2\sum_{NNN}{\textbf{S}_{i} \cdot \textbf{S}_{j}} +J_3\sum_{TNN}{\textbf{S}_{i} \cdot \textbf{S}_{j}}.
\label{hamiltonian}
\end{equation}
where sums run over nearest neighbor (NN), next-nearest neighbor (NNN) and third-nearest neighbor (TNN). Using linear spin wave theory (LSWT), we calculated single-magnon dispersion curves, shown as dotted lines in Fig. \ref{fig2}(a-b), and dynamical structure factor $S(\textbf{q},\omega)$, Fig. \ref{fig2}(a), for this model with best fit parameters J$_1$ = 1.45, J$_2$ = $-0.065$ and J$_3$ = 0.05 meV. Whilst the simplified model $(\ref{hamiltonian})$ reproduces the overall dispersion and intensities well, large differences at the B point, $\mathbf{Q}=(1/2, 1/2, 0)$, are apparent, including a clearly defined minimum in the measured dispersion which is absent in the calculation. We note that by increasing J$_2$/J$_1$ up to 1/6 or higher, the lower energy magnon mode can be flattened. However, analysis of the magnetic ground state points out that the classical 120$^{\circ}$ ordering can be stabilized without other coupling only if J$_2$/J$_1$ is less than 1/8 \cite{ivanov1993spin}.

An alternative explanation for the minimum at the B-point is the higher order (1/S) correction to LSWT which theoretical studies \cite{zheng2006anomalous,chernyshev2009spin} show can produce a pronounce downwards renormalization of the spin wave energies. However, Mourigal et al. \cite{mourigal2013dynamical} demonstrated that for the S=3/2 case, this renormalisation is just 8\% of the LSWT energies due to the large value of S. We previously observed similarly small 1/S corrections in our data for LuMnO$_3$ with S=2 \cite{JSOh2013MagnonDecay}. In CuCrO$_2$, though, this mechanism is insufficient to account for the observed minimum, motivating us to explore alternative spin Hamiltonians.

\subsection{B. Full Hamiltonian with magnon-phonon coupling}
\begin{figure}
\includegraphics[width=\columnwidth,clip]{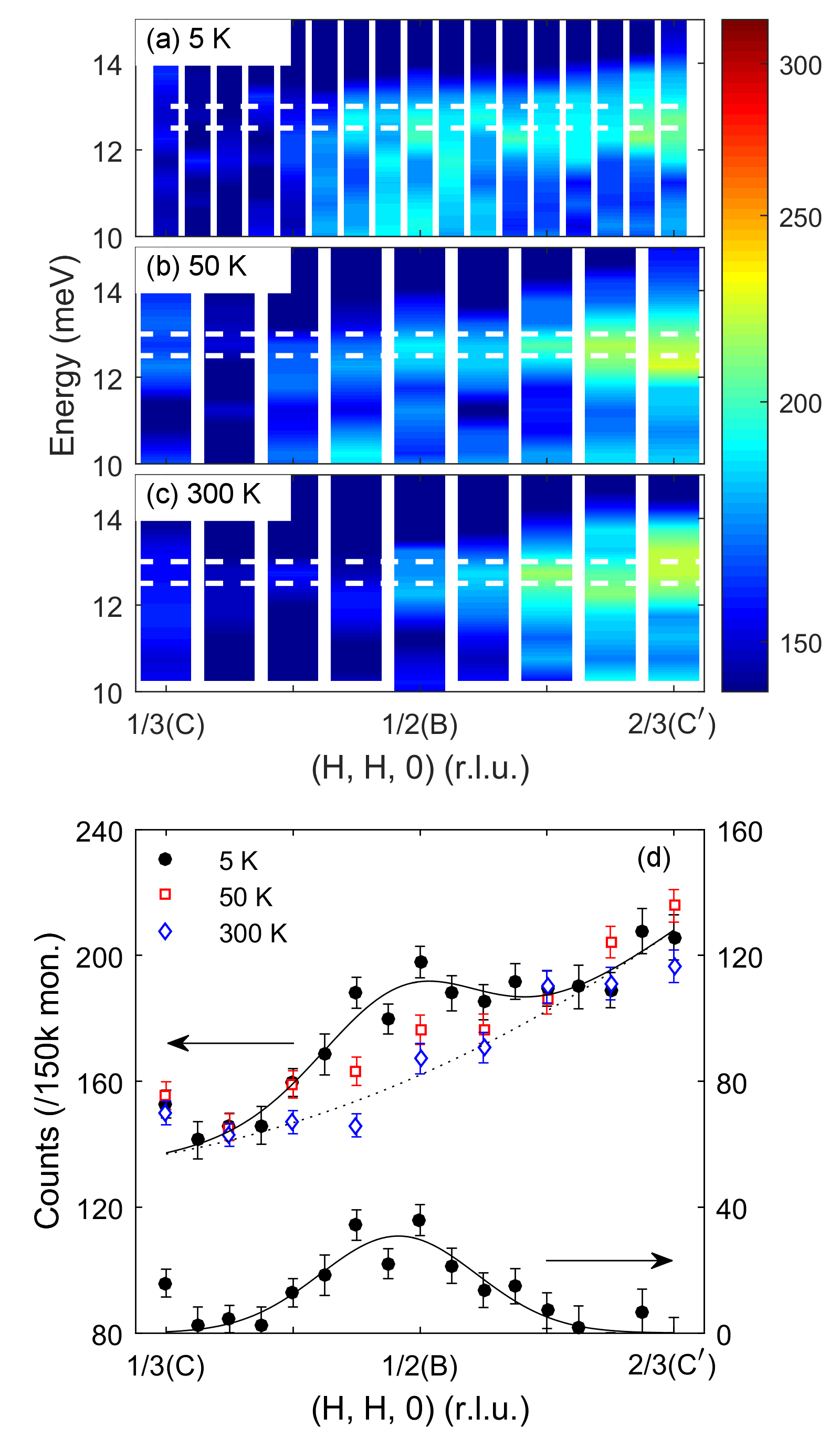}
\caption{\label{fig3}(Color online) (a-c) Contour plots along the C$-$B$-$C$^{\prime}$ path for the three indicated temperatures. (d) Temperature depenence of the neutron intensity integrated over a range of 12.5$-$13 meV, as marked by the white dashed lines in (a-c). The curve (solid line) at the bottom of (d) shows the phonon mode at 5 K with a $\textbf{q}^2$-contribution (dotted line) subtracted from the measured data points and fitted with a Gaussian function.}
\end{figure}

Besides the single-magnon spectrum, a flat phonon mode near 12.5 meV was identified in our data. Fig. \ref{fig3}(d) shows the integrated neutron intensity from 12.5 to 13 meV at three temperatures, as shown in Fig. \ref{fig3}(a-c). A phonon population (Bose) factor correction was applied to ensure that the data at 50 and 300 K could be compared directly to the 5 K data. No difference in this phonon intensity along the [H, H, 0] direction is observed above T$_N$ (50 and 300 K) and the intensity of the phonon mode has the usual $\textbf{q}^2$ dependence, as shown by the dotted line in Fig. \ref{fig3}(d). Strikingly though, there is a clear difference between the $\textbf{q}$-dependence of this phonon intensity at 5 K compared to the higher temperatures: a Gaussian-like signal with center at the B point (see the lower plot of Fig. \ref{fig3}(d)). Since this additional intensity emerges below the magnetic ordering temperature, it can be readily inferred that the magnetic order is strongly related to this phonon mode.

Many previous studies on the magneto-elastic properties in CuCrO$_2$ also give motivation to devise a new Hamiltonian that incorporates a magnon-phonon coupling. For example, X-ray diffraction and strain gauge measurement studies reported strong deformation of the triangular lattice plane below T$_N$ \cite{kimura2009Xdiff}. Also, ultrasound velocity measurements show the softening of the transverse acoustic phonon which accompanies the ferroelastic transition below T$_N$ \cite{aktas2013magnetoelastic}. More specifically, the temperature dependence of shifts in certain Raman-active peaks \cite{aktas2011raman} reveals that the frequency of the E$_g$ mode falls below T$_{N}$ in contrast with the A$_{1g}$ mode, which remains constant. These data implies that the coupling between the magnetic order and each phonon mode can vary, i.e. a specific lattice vibration mode can strongly interact with the magnetic structure.

It is worth mentioning that the direct exchange interaction between Cr$^{3+}$ ions is the dominant antiferromagnetic interaction in many Cr$^{3+}$ delafossite compounds \cite{doumerc1986magnetic,poienar2009Ndiff,kimura2009Xdiff}. Since the direct exchange is strongly affected by the distance between Cr$^{3+}$ ions, we can expect that the exchange parameter would be modulated by lattice vibrations. Therefore, with this well-known exchange-striction effect it is possible to modify the exchange interactions $J_{ij}$ \cite{pytte1974peierls,Jia2005mpcoupling,Bergman2006ESP,junghoon2007mpcoupling}:
\begin{equation}
J(\lvert i+u_{i}-j-u_{j}\rvert) \approx J_{1} + c_{mp} \hat{e}_{ji} \cdot (u_{i} - u_{j}).
\label{mpcoupling}
\end{equation}
where $\hat{e}_{ji}$ is a unit vector connecting Cr$^{3+}$ ions in equilibrium sites i and j, $u_{i}$ is atomic displacement from site i, and $c_{mp}$ is the magnitude of the first derivative of the exchange interaction with respects to the ionic displacements, thus coupling the magnetic order to lattice vibrations. Therefore, the full Hamiltonian can be expressed as following:
\begin{eqnarray}
{\cal H} = {\cal H}_{Heis} + \sum_{i} \left (\frac{p_{i}^{2}}{2m} + \frac{K}{2}{u_{i}^{2}} \right) \nonumber \\
+ c_{mp}\sum_{NN}{\hat{e}_{ji} \cdot (u_{i} - u_{j})\textbf{S}_{i} \cdot \textbf{S}_{j}}.
\label{hamiltonian2}
\end{eqnarray}
where the first term is the simple Heisenberg Hamiltonian given in Eq. $(\ref{hamiltonian})$ and the second term in the bracket is the usual phonon Hamiltonian.

The full Hamiltonian $(\ref{hamiltonian2})$ can be expressed in terms of single-magnon (single-phonon) creation, $\alpha_{\mathbf{k}}$ ($\beta_{\mathbf{k}}$), and anihilation, $\alpha^{\dagger}_{\mathbf{k}}$ ($\beta^{\dagger}_{\mathbf{k}}$), with the displacement $u_i$ being a single-operator term, whilst the spin $\mathbf{S}_i\cdot\mathbf{S}_j$ is a two-operator term in the linear approximation. Thus the exchange-striction coupling term is cubic in the boson operators, e.g. $\alpha_{\mathbf{k}}\alpha^{\dagger}_{\mathbf{k}}\beta_{\mathbf{k}}$, and so should be weak. However, for non-collinear magnetic structures and going beyond LSWT, $\mathbf{S}_i\cdot\mathbf{S}_j$ also yields one-operator and three-operator terms \cite{chernyshev2009spin}. The one-magnon term can thus couple linearly with the one-phonon term, resulting in terms like $\alpha_{\mathbf{k}}\beta^{\dagger}_{\mathbf{k}}$ and hence a stronger hybridization between phonon and magnon modes.

In principle, a magnon can couple with all of the phonon modes. However, as discussed above, coupling strength between magnons and phonons varies depending on the particular magnon or phonon modes involved. In this case, we observed that the intensity of the phonon at 12.5 meV is greatly enhanced below T$_N$ so it is reasonable to use this mode in the calculation (for computational reasons we do not consider the full phonon dispersion but only couple one Einstein phonon mode to the previously calculated magnon spectrum). Then, there are two reasons why the magnon-phonon coupling strength can be expected to be strongest around this phonon mode: First, the coupling between quasiparticles is expected to be large when the two modes are close in the momentum-energy space \cite{stone2006quasiparticle,zhitomirsky2013colloquium}. Second, due to the fact that the magnon-phonon coupling is proportional to inverse square-root of energy of phonons \cite{white1965diagonalization, jsoh2016}, the contribution of a high energy phonon may be small even if all of phonon modes have the same coupling constant.

Using this full Hamiltonian, the following best fit parameters were obtained: J$_1$ = 1.57, J$_2$ = $-0.045$ and J$_3$ = 0.145 meV, $c_{mp}$ = 16.8 meV/\AA. The resulting dispersion is shown as the solid line in Fig. \ref{fig2}(b-c) and the calculated $S(\textbf{q},\omega)$ in Fig. \ref{fig2}(c). The model now contains both magnon and coupled magnon-phonon mixed character and reproduces the 5 meV minima at the B point with the correct intensity, as well as the high energy signal at 12.5 meV. We note that the calculated $S(\textbf{q},\omega)$ doesn’t include the $\textbf{q}^2$-contribution since we didn’t consider the nuclear structure factor related to lattice vibrations in the model Hamiltonian $(\ref{hamiltonian2})$. The good agreement between the model calculations and measurements thus strongly suggests that the observed deviations from LSWT originates, at least partly, in the magnon-phonon coupling.

\begin{figure}[!htb]
\includegraphics[width=\columnwidth,clip]{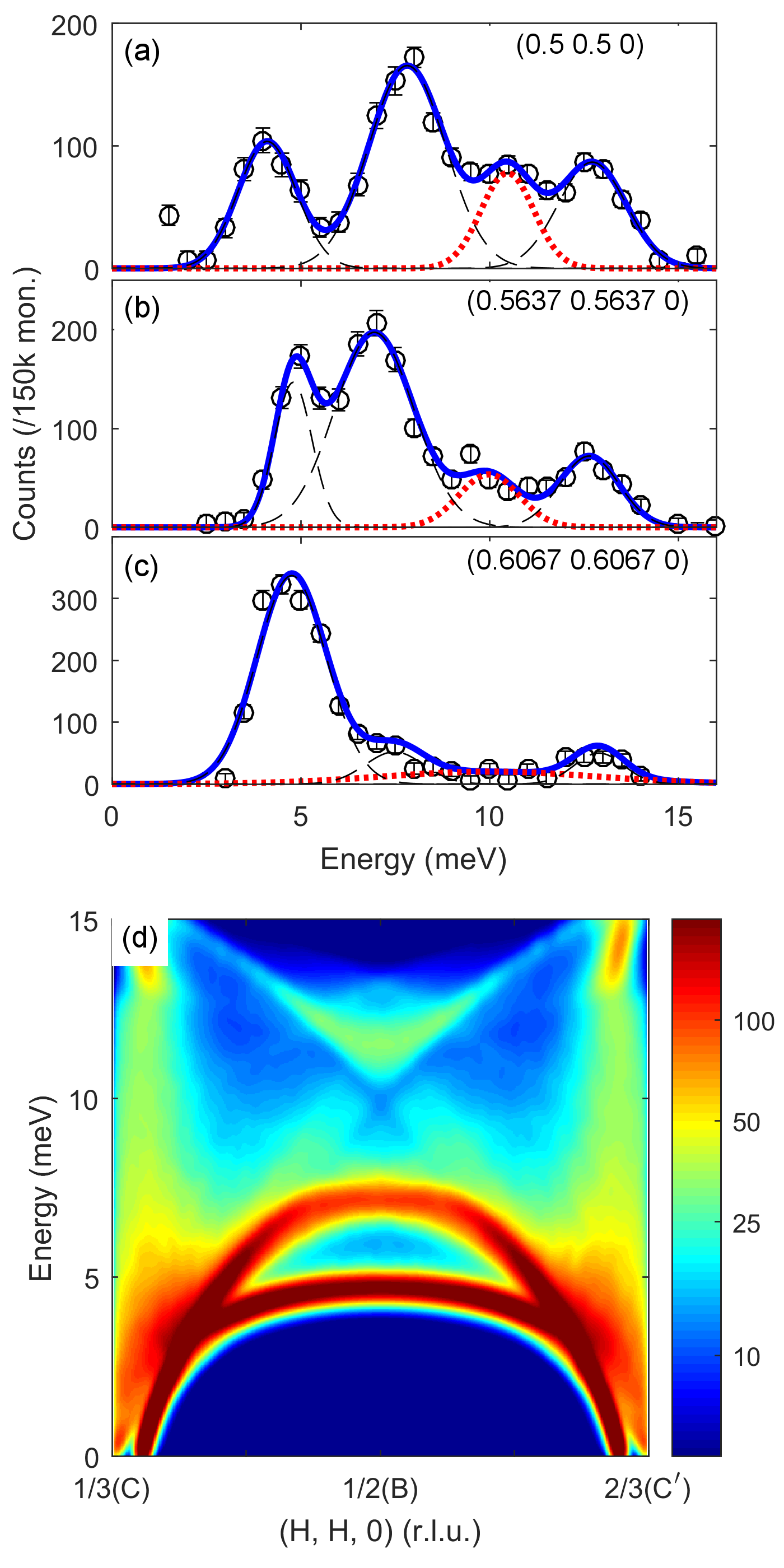}
\caption{\label{fig4}(Color online) Constant $\textbf{q}$-cut with Gaussian fitting at different $\textbf{q}$ = (a) $(0.5, 0.5, 0)$, (b) $(0.5637, 0.5637, 0)$ and (c) $(0.6067, 0.6067, 0)$. Background was subtracted off in all plots. Dashed lines denotes the individual Gaussian peak of single magnon and phonon (12.5 meV). Dotted lines near 10 meV are intensity of the continuum excitation, while the solid line shows the summation of all peaks. (d) Calculated two-magnon continuum using the minimal Heisenberg model in Eq. $(\ref{hamiltonian})$.}
\end{figure}

One remarkable deviation between our model calculations and the data comes from the upward shift of the phonon mode near the B point. This is an example of the phenomena of level repulsion: for quasiparticles with coupling between them, the dispersions of the quasiparticles repel each other by bending along the opposite directions, or they open a gap to avoid crossing. In CuCrO$_2$, due to the large coupling between magnon and phonon modes, these two modes repel each other. However, the degree of bending in the 12.5 meV phonon mode as shown near B point in our model (Fig. \ref{fig2}(c)) is bigger than in our measurement. It is an exaggerated feature of our simplified model since it assumes just one phonon mode, which forces the revel repulsion to be of the same degree as the magnon mode. Further studies with theoretical calculations of the full phonon dispersion and the density of states are needed to estimate the actual contributions of all phonon modes accurately.

\section{IV. Discussion}
To gain further insight into the magnitude of the magnon-phonon coupling constant, we can use high-pressure experimental data \cite{aoyama2013pressureEffect,garg2014pressureRaman} from the CuCrO$_2$ literature to estimate $c_{mp}$. In case of a spin-lattice coupled compound, when pressure is applied on the material, reduced atomic distances bring about the modulation of the exchange interaction, which results subsequently in the modulation of T$_N$. So it is possible to estimate the exchange-striction coefficient $\alpha$ and the magnon-phonon coupling constant $c_{mp}$ using the following formula:
\begin{equation}
\alpha = \frac{c_{mp}}{J_1} = \frac{1}{T_N} \frac{\Delta T_N / \Delta p} {\Delta a / \Delta p}
\label{couplingconstantestimation}
\end{equation}

Using the data found in Ref. \cite{aoyama2013pressureEffect,garg2014pressureRaman}, we estimate the values of $\alpha \approx$ 10 \AA$^{-1}$ and $c_{mp}$ = 17.21(46) meV/\AA. These estimations are in good agreement with our best fit parameter ($c_{mp}$ = 16.8 meV/\AA), which shows that our assumptions are legitimate. It is worth mentioning that the exchange-striction coefficient $\alpha$ of Cr$^{3+}$ which we have determined is larger compared with the cases of other ions such as Cu$^{2+}$ ($\approx$ 7 \AA$^{-1}$) \cite{aronson1991pressure}. Considering that $\alpha$ is equal to the exponent of the power law of the exchange interaction $J \propto 1/r{^n}$, such a large value of $\alpha$ is consistent with the fact that the exchange interaction between Cr$^{3+}$ is dominated by the direct exchange between the Cr$^{3+}$ orbitals as discussed above.

The last feature worthy of mentioning in our inelastic neutron spectra is the continuum within energy transfer $\approx$ 10.5$-$11 meV at the B point. Interestingly, it is located at $\omega = 7JS$, which is exactly the same position as that of the calculated two-magnon continuum in ref. \cite{mourigal2013dynamical} for the S=3/2 case. To investigate this in further detail, we plot energy scans at three different $\textbf{q}$ points along the [H, H, 0] direction, as shown in Fig. \ref{fig4}(a-c). Each plot was fitted with several Gaussian functions after the background signals were removed. For the B point at (0.5, 0.5, 0), the continuum excitation (see dotted red line) has strong intensity and a sharp peak shape comparable with other signals from the single quasiparticle spectrum. With increasing H in the [H, H, 0] direction, the signal is dispersed and weakened.

To quantitatively compare these observed continuum peaks with theory, we calculated the dynamical structure factor using the nonlinear spin wave theory with 1/S formalism on the interacting spin waves. The numerical integration of the magnon self-energy was implemented using Eq. (13) in ref. \cite{mourigal2013dynamical}. Randomly chosen $6 \times 10^{6}$ $\textbf{q}$ points within first Brillouin zone were used in a Monte Carlo simulation to compute the self-energy of the given $\textbf{k}$ point. The tendency shown in Fig. \ref{fig4}(a-c) is consistent with our calculated two-magnon dispersion in Fig. \ref{fig4}(d). However, the calculated intensity of this continuum is not so strong as it is in our experimental observations. As a possible explanation, we can consider a new decay channel related to the magnons and phonons. Due to the interaction between magnons and phonons, there are new cubic terms such as $\alpha_{\textbf{k}}$$\beta^{\dagger}_{\textbf{k}+\textbf{q}}$$\beta_{\textbf{q}}$, making the intensity of continuum stronger. Note that we used the minimal Heisenberg model of Eq. $(\ref{hamiltonian})$ for the two-magnon calculation since it is difficult to calculate the 1/S correction terms with inclusion of phonon operators.

\section{V. Conclusion}
In conclusion, we have mapped out the INS spectra in the triangular antiferromagnet CuCrO$_2$. The overall magnon dispersion is explained by our simplified model with a single dispersionless phonon mode and the non-vanishing linear coupling between magnons and phonons originating from the non-collinear magnetic structure. This large interaction between the magnons and phonons leads to a minimum in the magnon dispersion and corresponding maximum in the phonon spectrum as a result of level repulsion. Continuum excitations due to the interactions between quasiparticles have tendencies consistent with our calculations. The concept of magnon-phonon coupling used in this paper could be applied more generally in other magnetic materials with non-collinear magnetic order. The experimental results and the analysis reported here constitutes another nontrivial aspect of the noncollinear 120$^{\circ}$ magnetic structure of the triangular lattice.

\section{Acknowledgements}
The work at the IBS CCES was supported by the research programme of Institute for Basic Science (IBS-R009-G1).

\begin{thebibliography}{28}%
\makeatletter
\providecommand \@ifxundefined [1]{%
 \@ifx{#1\undefined}
}%
\providecommand \@ifnum [1]{%
 \ifnum #1\expandafter \@firstoftwo
 \else \expandafter \@secondoftwo
 \fi
}%
\providecommand \@ifx [1]{%
 \ifx #1\expandafter \@firstoftwo
 \else \expandafter \@secondoftwo
 \fi
}%
\providecommand \natexlab [1]{#1}%
\providecommand \enquote  [1]{``#1''}%
\providecommand \bibnamefont  [1]{#1}%
\providecommand \bibfnamefont [1]{#1}%
\providecommand \citenamefont [1]{#1}%
\providecommand \href@noop [0]{\@secondoftwo}%
\providecommand \href [0]{\begingroup \@sanitize@url \@href}%
\providecommand \@href[1]{\@@startlink{#1}\@@href}%
\providecommand \@@href[1]{\endgroup#1\@@endlink}%
\providecommand \@sanitize@url [0]{\catcode `\\12\catcode `\$12\catcode
  `\&12\catcode `\#12\catcode `\^12\catcode `\_12\catcode `\%12\relax}%
\providecommand \@@startlink[1]{}%
\providecommand \@@endlink[0]{}%
\providecommand \url  [0]{\begingroup\@sanitize@url \@url }%
\providecommand \@url [1]{\endgroup\@href {#1}{\urlprefix }}%
\providecommand \urlprefix  [0]{URL }%
\providecommand \Eprint [0]{\href }%
\providecommand \doibase [0]{http://dx.doi.org/}%
\providecommand \selectlanguage [0]{\@gobble}%
\providecommand \bibinfo  [0]{\@secondoftwo}%
\providecommand \bibfield  [0]{\@secondoftwo}%
\providecommand \translation [1]{[#1]}%
\providecommand \BibitemOpen [0]{}%
\providecommand \bibitemStop [0]{}%
\providecommand \bibitemNoStop [0]{.\EOS\space}%
\providecommand \EOS [0]{\spacefactor3000\relax}%
\providecommand \BibitemShut  [1]{\csname bibitem#1\endcsname}%
\let\auto@bib@innerbib\@empty
\bibitem [{\citenamefont {Seki}\ \emph {et~al.}(2008)\citenamefont {Seki},
  \citenamefont {Onose},\ and\ \citenamefont {Tokura}}]{seki2008spin}%
  \BibitemOpen
  \bibfield  {author} {\bibinfo {author} {\bibfnamefont {S.}~\bibnamefont
  {Seki}}, \bibinfo {author} {\bibfnamefont {Y.}~\bibnamefont {Onose}}, \ and\
  \bibinfo {author} {\bibfnamefont {Y.}~\bibnamefont {Tokura}},\ }\href@noop {}
  {\bibfield  {journal} {\bibinfo  {journal} {Phys. Rev. Lett.}\ }\textbf
  {\bibinfo {volume} {101}},\ \bibinfo {pages} {067204} (\bibinfo {year}
  {2008})}\BibitemShut {NoStop}%
\bibitem [{\citenamefont {Poienar}\ \emph {et~al.}(2009)\citenamefont
  {Poienar}, \citenamefont {Damay}, \citenamefont {Martin}, \citenamefont
  {Hardy}, \citenamefont {Maignan},\ and\ \citenamefont
  {Andr{\'e}}}]{poienar2009Ndiff}%
  \BibitemOpen
  \bibfield  {author} {\bibinfo {author} {\bibfnamefont {M.}~\bibnamefont
  {Poienar}}, \bibinfo {author} {\bibfnamefont {F.}~\bibnamefont {Damay}},
  \bibinfo {author} {\bibfnamefont {C.}~\bibnamefont {Martin}}, \bibinfo
  {author} {\bibfnamefont {V.}~\bibnamefont {Hardy}}, \bibinfo {author}
  {\bibfnamefont {A.}~\bibnamefont {Maignan}}, \ and\ \bibinfo {author}
  {\bibfnamefont {G.}~\bibnamefont {Andr{\'e}}},\ }\href@noop {} {\bibfield
  {journal} {\bibinfo  {journal} {Phys. Rev. B}\ }\textbf {\bibinfo {volume}
  {79}},\ \bibinfo {pages} {014412} (\bibinfo {year} {2009})}\BibitemShut
  {NoStop}%
\bibitem [{\citenamefont {Poienar}\ \emph {et~al.}(2010)\citenamefont
  {Poienar}, \citenamefont {Damay}, \citenamefont {Martin}, \citenamefont
  {Robert},\ and\ \citenamefont {Petit}}]{Poienar2010INS}%
  \BibitemOpen
  \bibfield  {author} {\bibinfo {author} {\bibfnamefont {M.}~\bibnamefont
  {Poienar}}, \bibinfo {author} {\bibfnamefont {F.}~\bibnamefont {Damay}},
  \bibinfo {author} {\bibfnamefont {C.}~\bibnamefont {Martin}}, \bibinfo
  {author} {\bibfnamefont {J.}~\bibnamefont {Robert}}, \ and\ \bibinfo {author}
  {\bibfnamefont {S.}~\bibnamefont {Petit}},\ }\href {\doibase
  10.1103/PhysRevB.81.104411} {\bibfield  {journal} {\bibinfo  {journal} {Phys.
  Rev. B}\ }\textbf {\bibinfo {volume} {81}},\ \bibinfo {pages} {104411}
  (\bibinfo {year} {2010})}\BibitemShut {NoStop}%
\bibitem [{\citenamefont {Frontzek}\ \emph
  {et~al.}(2011{\natexlab{a}})\citenamefont {Frontzek}, \citenamefont
  {Haraldsen}, \citenamefont {Podlesnyak}, \citenamefont {Matsuda},
  \citenamefont {Christianson}, \citenamefont {Fishman}, \citenamefont {Sefat},
  \citenamefont {Qiu}, \citenamefont {Copley}, \citenamefont {Barilo},
  \citenamefont {Shiryaev},\ and\ \citenamefont {Ehlers}}]{FrontzekINS2011}%
  \BibitemOpen
  \bibfield  {author} {\bibinfo {author} {\bibfnamefont {M.}~\bibnamefont
  {Frontzek}}, \bibinfo {author} {\bibfnamefont {J.~T.}\ \bibnamefont
  {Haraldsen}}, \bibinfo {author} {\bibfnamefont {A.}~\bibnamefont
  {Podlesnyak}}, \bibinfo {author} {\bibfnamefont {M.}~\bibnamefont {Matsuda}},
  \bibinfo {author} {\bibfnamefont {A.~D.}\ \bibnamefont {Christianson}},
  \bibinfo {author} {\bibfnamefont {R.~S.}\ \bibnamefont {Fishman}}, \bibinfo
  {author} {\bibfnamefont {A.~S.}\ \bibnamefont {Sefat}}, \bibinfo {author}
  {\bibfnamefont {Y.}~\bibnamefont {Qiu}}, \bibinfo {author} {\bibfnamefont
  {J.~R.~D.}\ \bibnamefont {Copley}}, \bibinfo {author} {\bibfnamefont
  {S.}~\bibnamefont {Barilo}}, \bibinfo {author} {\bibfnamefont {S.~V.}\
  \bibnamefont {Shiryaev}}, \ and\ \bibinfo {author} {\bibfnamefont
  {G.}~\bibnamefont {Ehlers}},\ }\href {\doibase 10.1103/PhysRevB.84.094448}
  {\bibfield  {journal} {\bibinfo  {journal} {Phys. Rev. B}\ }\textbf {\bibinfo
  {volume} {84}},\ \bibinfo {pages} {094448} (\bibinfo {year}
  {2011}{\natexlab{a}})}\BibitemShut {NoStop}%
\bibitem [{\citenamefont {Kajimoto}\ \emph {et~al.}(2015)\citenamefont
  {Kajimoto}, \citenamefont {Tomiyasu}, \citenamefont {Nakajima}, \citenamefont
  {Ohira-Kawamura}, \citenamefont {Inamura},\ and\ \citenamefont
  {Okuda}}]{Kajimoto2015INSdiffusion}%
  \BibitemOpen
  \bibfield  {author} {\bibinfo {author} {\bibfnamefont {R.}~\bibnamefont
  {Kajimoto}}, \bibinfo {author} {\bibfnamefont {K.}~\bibnamefont {Tomiyasu}},
  \bibinfo {author} {\bibfnamefont {K.}~\bibnamefont {Nakajima}}, \bibinfo
  {author} {\bibfnamefont {S.}~\bibnamefont {Ohira-Kawamura}}, \bibinfo
  {author} {\bibfnamefont {Y.}~\bibnamefont {Inamura}}, \ and\ \bibinfo
  {author} {\bibfnamefont {T.}~\bibnamefont {Okuda}},\ }\href@noop {}
  {\bibfield  {journal} {\bibinfo  {journal} {J. Phys. Soc. Jpn.}\ }\textbf
  {\bibinfo {volume} {84}},\ \bibinfo {pages} {074708} (\bibinfo {year}
  {2015})}\BibitemShut {NoStop}%
\bibitem [{\citenamefont {Y{\'e}}\ \emph {et~al.}(1994)\citenamefont {Y{\'e}},
  \citenamefont {Crottaz}, \citenamefont {Vaudano}, \citenamefont {Kubel},
  \citenamefont {Tissot},\ and\ \citenamefont {Schmid}}]{Ye1994Growth}%
  \BibitemOpen
  \bibfield  {author} {\bibinfo {author} {\bibfnamefont {Z.-G.}\ \bibnamefont
  {Y{\'e}}}, \bibinfo {author} {\bibfnamefont {O.}~\bibnamefont {Crottaz}},
  \bibinfo {author} {\bibfnamefont {F.}~\bibnamefont {Vaudano}}, \bibinfo
  {author} {\bibfnamefont {F.}~\bibnamefont {Kubel}}, \bibinfo {author}
  {\bibfnamefont {P.}~\bibnamefont {Tissot}}, \ and\ \bibinfo {author}
  {\bibfnamefont {H.}~\bibnamefont {Schmid}},\ }\href@noop {} {\bibfield
  {journal} {\bibinfo  {journal} {Ferroelectrics}\ }\textbf {\bibinfo {volume}
  {162}},\ \bibinfo {pages} {103} (\bibinfo {year} {1994})}\BibitemShut
  {NoStop}%
\bibitem [{\citenamefont {Danilkin}\ and\ \citenamefont
  {Yethiraj}(2009)}]{danilkin2009taipan}%
  \BibitemOpen
  \bibfield  {author} {\bibinfo {author} {\bibfnamefont {S.}~\bibnamefont
  {Danilkin}}\ and\ \bibinfo {author} {\bibfnamefont {M.}~\bibnamefont
  {Yethiraj}},\ }\href@noop {} {\bibfield  {journal} {\bibinfo  {journal}
  {Neutron News}\ }\textbf {\bibinfo {volume} {20}},\ \bibinfo {pages} {37}
  (\bibinfo {year} {2009})}\BibitemShut {NoStop}%
\bibitem [{\citenamefont {Frontzek}\ \emph
  {et~al.}(2011{\natexlab{b}})\citenamefont {Frontzek}, \citenamefont {Ehlers},
  \citenamefont {Podlesnyak}, \citenamefont {Cao}, \citenamefont {Matsuda},
  \citenamefont {Zaharko}, \citenamefont {Aliouane}, \citenamefont {Barilo},\
  and\ \citenamefont {Shiryaev}}]{frontzek2011NDiff}%
  \BibitemOpen
  \bibfield  {author} {\bibinfo {author} {\bibfnamefont {M.}~\bibnamefont
  {Frontzek}}, \bibinfo {author} {\bibfnamefont {G.}~\bibnamefont {Ehlers}},
  \bibinfo {author} {\bibfnamefont {A.}~\bibnamefont {Podlesnyak}}, \bibinfo
  {author} {\bibfnamefont {H.}~\bibnamefont {Cao}}, \bibinfo {author}
  {\bibfnamefont {M.}~\bibnamefont {Matsuda}}, \bibinfo {author} {\bibfnamefont
  {O.}~\bibnamefont {Zaharko}}, \bibinfo {author} {\bibfnamefont
  {N.}~\bibnamefont {Aliouane}}, \bibinfo {author} {\bibfnamefont
  {S.}~\bibnamefont {Barilo}}, \ and\ \bibinfo {author} {\bibfnamefont
  {S.}~\bibnamefont {Shiryaev}},\ }\href@noop {} {\bibfield  {journal}
  {\bibinfo  {journal} {J. Phys: Condens. Matter}\ }\textbf {\bibinfo {volume}
  {24}},\ \bibinfo {pages} {016004} (\bibinfo {year}
  {2011}{\natexlab{b}})}\BibitemShut {NoStop}%
\bibitem [{\citenamefont {Ivanov}(1993)}]{ivanov1993spin}%
  \BibitemOpen
  \bibfield  {author} {\bibinfo {author} {\bibfnamefont {N.}~\bibnamefont
  {Ivanov}},\ }\href@noop {} {\bibfield  {journal} {\bibinfo  {journal} {Phys.
  Rev. B}\ }\textbf {\bibinfo {volume} {47}},\ \bibinfo {pages} {9105}
  (\bibinfo {year} {1993})}\BibitemShut {NoStop}%
\bibitem [{\citenamefont {Zheng}\ \emph {et~al.}(2006)\citenamefont {Zheng},
  \citenamefont {Fj{\ae}restad}, \citenamefont {Singh}, \citenamefont
  {McKenzie},\ and\ \citenamefont {Coldea}}]{zheng2006anomalous}%
  \BibitemOpen
  \bibfield  {author} {\bibinfo {author} {\bibfnamefont {W.}~\bibnamefont
  {Zheng}}, \bibinfo {author} {\bibfnamefont {J.~O.}\ \bibnamefont
  {Fj{\ae}restad}}, \bibinfo {author} {\bibfnamefont {R.~R.}\ \bibnamefont
  {Singh}}, \bibinfo {author} {\bibfnamefont {R.~H.}\ \bibnamefont {McKenzie}},
  \ and\ \bibinfo {author} {\bibfnamefont {R.}~\bibnamefont {Coldea}},\
  }\href@noop {} {\bibfield  {journal} {\bibinfo  {journal} {Phys. Rev. Lett.}\
  }\textbf {\bibinfo {volume} {96}},\ \bibinfo {pages} {057201} (\bibinfo
  {year} {2006})}\BibitemShut {NoStop}%
\bibitem [{\citenamefont {Chernyshev}\ and\ \citenamefont
  {Zhitomirsky}(2009)}]{chernyshev2009spin}%
  \BibitemOpen
  \bibfield  {author} {\bibinfo {author} {\bibfnamefont {A.}~\bibnamefont
  {Chernyshev}}\ and\ \bibinfo {author} {\bibfnamefont {M.}~\bibnamefont
  {Zhitomirsky}},\ }\href@noop {} {\bibfield  {journal} {\bibinfo  {journal}
  {Phys. Rev. B}\ }\textbf {\bibinfo {volume} {79}},\ \bibinfo {pages} {144416}
  (\bibinfo {year} {2009})}\BibitemShut {NoStop}%
\bibitem [{\citenamefont {Mourigal}\ \emph {et~al.}(2013)\citenamefont
  {Mourigal}, \citenamefont {Fuhrman}, \citenamefont {Chernyshev},\ and\
  \citenamefont {Zhitomirsky}}]{mourigal2013dynamical}%
  \BibitemOpen
  \bibfield  {author} {\bibinfo {author} {\bibfnamefont {M.}~\bibnamefont
  {Mourigal}}, \bibinfo {author} {\bibfnamefont {W.}~\bibnamefont {Fuhrman}},
  \bibinfo {author} {\bibfnamefont {A.}~\bibnamefont {Chernyshev}}, \ and\
  \bibinfo {author} {\bibfnamefont {M.}~\bibnamefont {Zhitomirsky}},\
  }\href@noop {} {\bibfield  {journal} {\bibinfo  {journal} {Phys. Rev. B}\
  }\textbf {\bibinfo {volume} {88}},\ \bibinfo {pages} {094407} (\bibinfo
  {year} {2013})}\BibitemShut {NoStop}%
\bibitem [{\citenamefont {Oh}\ \emph {et~al.}(2013)\citenamefont {Oh},
  \citenamefont {Le}, \citenamefont {Jeong}, \citenamefont {Lee}, \citenamefont
  {Woo}, \citenamefont {Song}, \citenamefont {Perring}, \citenamefont {Buyers},
  \citenamefont {Cheong},\ and\ \citenamefont {Park}}]{JSOh2013MagnonDecay}%
  \BibitemOpen
  \bibfield  {author} {\bibinfo {author} {\bibfnamefont {J.}~\bibnamefont
  {Oh}}, \bibinfo {author} {\bibfnamefont {M.~D.}\ \bibnamefont {Le}}, \bibinfo
  {author} {\bibfnamefont {J.}~\bibnamefont {Jeong}}, \bibinfo {author}
  {\bibfnamefont {J.-H.}\ \bibnamefont {Lee}}, \bibinfo {author} {\bibfnamefont
  {H.}~\bibnamefont {Woo}}, \bibinfo {author} {\bibfnamefont {W.-Y.}\
  \bibnamefont {Song}}, \bibinfo {author} {\bibfnamefont {T.~G.}\ \bibnamefont
  {Perring}}, \bibinfo {author} {\bibfnamefont {W.~J.~L.}\ \bibnamefont
  {Buyers}}, \bibinfo {author} {\bibfnamefont {S.-W.}\ \bibnamefont {Cheong}},
  \ and\ \bibinfo {author} {\bibfnamefont {J.-G.}\ \bibnamefont {Park}},\
  }\href {\doibase 10.1103/PhysRevLett.111.257202} {\bibfield  {journal}
  {\bibinfo  {journal} {Phys. Rev. Lett.}\ }\textbf {\bibinfo {volume} {111}},\
  \bibinfo {pages} {257202} (\bibinfo {year} {2013})}\BibitemShut {NoStop}%
\bibitem [{\citenamefont {Kimura}\ \emph {et~al.}(2009)\citenamefont {Kimura},
  \citenamefont {Otani}, \citenamefont {Nakamura}, \citenamefont
  {Wakabayashi},\ and\ \citenamefont {Kimura}}]{kimura2009Xdiff}%
  \BibitemOpen
  \bibfield  {author} {\bibinfo {author} {\bibfnamefont {K.}~\bibnamefont
  {Kimura}}, \bibinfo {author} {\bibfnamefont {T.}~\bibnamefont {Otani}},
  \bibinfo {author} {\bibfnamefont {H.}~\bibnamefont {Nakamura}}, \bibinfo
  {author} {\bibfnamefont {Y.}~\bibnamefont {Wakabayashi}}, \ and\ \bibinfo
  {author} {\bibfnamefont {T.}~\bibnamefont {Kimura}},\ }\href@noop {}
  {\bibfield  {journal} {\bibinfo  {journal} {J. Phys. Soc. Jpn.}\ }\textbf
  {\bibinfo {volume} {78}},\ \bibinfo {pages} {113710} (\bibinfo {year}
  {2009})}\BibitemShut {NoStop}%
\bibitem [{\citenamefont {Aktas}\ \emph {et~al.}(2013)\citenamefont {Aktas},
  \citenamefont {Quirion}, \citenamefont {Otani},\ and\ \citenamefont
  {Kimura}}]{aktas2013magnetoelastic}%
  \BibitemOpen
  \bibfield  {author} {\bibinfo {author} {\bibfnamefont {O.}~\bibnamefont
  {Aktas}}, \bibinfo {author} {\bibfnamefont {G.}~\bibnamefont {Quirion}},
  \bibinfo {author} {\bibfnamefont {T.}~\bibnamefont {Otani}}, \ and\ \bibinfo
  {author} {\bibfnamefont {T.}~\bibnamefont {Kimura}},\ }\href@noop {}
  {\bibfield  {journal} {\bibinfo  {journal} {Phys. Rev. B}\ }\textbf {\bibinfo
  {volume} {88}},\ \bibinfo {pages} {224104} (\bibinfo {year}
  {2013})}\BibitemShut {NoStop}%
\bibitem [{\citenamefont {Aktas}\ \emph {et~al.}(2011)\citenamefont {Aktas},
  \citenamefont {Truong}, \citenamefont {Otani}, \citenamefont {Balakrishnan},
  \citenamefont {Clouter}, \citenamefont {Kimura},\ and\ \citenamefont
  {Quirion}}]{aktas2011raman}%
  \BibitemOpen
  \bibfield  {author} {\bibinfo {author} {\bibfnamefont {O.}~\bibnamefont
  {Aktas}}, \bibinfo {author} {\bibfnamefont {K.~D.}\ \bibnamefont {Truong}},
  \bibinfo {author} {\bibfnamefont {T.}~\bibnamefont {Otani}}, \bibinfo
  {author} {\bibfnamefont {G.}~\bibnamefont {Balakrishnan}}, \bibinfo {author}
  {\bibfnamefont {M.~J.}\ \bibnamefont {Clouter}}, \bibinfo {author}
  {\bibfnamefont {T.}~\bibnamefont {Kimura}}, \ and\ \bibinfo {author}
  {\bibfnamefont {G.}~\bibnamefont {Quirion}},\ }\href@noop {} {\bibfield
  {journal} {\bibinfo  {journal} {J. Phys: Condens. Matter}\ }\textbf {\bibinfo
  {volume} {24}},\ \bibinfo {pages} {036003} (\bibinfo {year}
  {2011})}\BibitemShut {NoStop}%
\bibitem [{\citenamefont {Doumerc}\ \emph {et~al.}(1986)\citenamefont
  {Doumerc}, \citenamefont {Wichainchai}, \citenamefont {Ammar}, \citenamefont
  {Pouchard},\ and\ \citenamefont {Hagenmuller}}]{doumerc1986magnetic}%
  \BibitemOpen
  \bibfield  {author} {\bibinfo {author} {\bibfnamefont {J.-P.}\ \bibnamefont
  {Doumerc}}, \bibinfo {author} {\bibfnamefont {A.}~\bibnamefont
  {Wichainchai}}, \bibinfo {author} {\bibfnamefont {A.}~\bibnamefont {Ammar}},
  \bibinfo {author} {\bibfnamefont {M.}~\bibnamefont {Pouchard}}, \ and\
  \bibinfo {author} {\bibfnamefont {P.}~\bibnamefont {Hagenmuller}},\
  }\href@noop {} {\bibfield  {journal} {\bibinfo  {journal} {Mater. Res.
  Bull.}\ }\textbf {\bibinfo {volume} {21}},\ \bibinfo {pages} {745} (\bibinfo
  {year} {1986})}\BibitemShut {NoStop}%
\bibitem [{\citenamefont {Pytte}(1974)}]{pytte1974peierls}%
  \BibitemOpen
  \bibfield  {author} {\bibinfo {author} {\bibfnamefont {E.}~\bibnamefont
  {Pytte}},\ }\href@noop {} {\bibfield  {journal} {\bibinfo  {journal} {Phys.
  Rev. B}\ }\textbf {\bibinfo {volume} {10}},\ \bibinfo {pages} {4637}
  (\bibinfo {year} {1974})}\BibitemShut {NoStop}%
\bibitem [{\citenamefont {Jia}\ \emph {et~al.}(2005)\citenamefont {Jia},
  \citenamefont {Nam}, \citenamefont {Kim},\ and\ \citenamefont
  {Han}}]{Jia2005mpcoupling}%
  \BibitemOpen
  \bibfield  {author} {\bibinfo {author} {\bibfnamefont {C.}~\bibnamefont
  {Jia}}, \bibinfo {author} {\bibfnamefont {J.~H.}\ \bibnamefont {Nam}},
  \bibinfo {author} {\bibfnamefont {J.~S.}\ \bibnamefont {Kim}}, \ and\
  \bibinfo {author} {\bibfnamefont {J.~H.}\ \bibnamefont {Han}},\ }\href
  {\doibase 10.1103/PhysRevB.71.212406} {\bibfield  {journal} {\bibinfo
  {journal} {Phys. Rev. B}\ }\textbf {\bibinfo {volume} {71}},\ \bibinfo
  {pages} {212406} (\bibinfo {year} {2005})}\BibitemShut {NoStop}%
\bibitem [{\citenamefont {Bergman}\ \emph {et~al.}(2006)\citenamefont
  {Bergman}, \citenamefont {Shindou}, \citenamefont {Fiete},\ and\
  \citenamefont {Balents}}]{Bergman2006ESP}%
  \BibitemOpen
  \bibfield  {author} {\bibinfo {author} {\bibfnamefont {D.~L.}\ \bibnamefont
  {Bergman}}, \bibinfo {author} {\bibfnamefont {R.}~\bibnamefont {Shindou}},
  \bibinfo {author} {\bibfnamefont {G.~A.}\ \bibnamefont {Fiete}}, \ and\
  \bibinfo {author} {\bibfnamefont {L.}~\bibnamefont {Balents}},\ }\href
  {\doibase 10.1103/PhysRevB.74.134409} {\bibfield  {journal} {\bibinfo
  {journal} {Phys. Rev. B}\ }\textbf {\bibinfo {volume} {74}},\ \bibinfo
  {pages} {134409} (\bibinfo {year} {2006})}\BibitemShut {NoStop}%
\bibitem [{\citenamefont {Kim}\ and\ \citenamefont
  {Han}(2007)}]{junghoon2007mpcoupling}%
  \BibitemOpen
  \bibfield  {author} {\bibinfo {author} {\bibfnamefont {J.~H.}\ \bibnamefont
  {Kim}}\ and\ \bibinfo {author} {\bibfnamefont {J.~H.}\ \bibnamefont {Han}},\
  }\href@noop {} {\bibfield  {journal} {\bibinfo  {journal} {Phys. Rev. B}\
  }\textbf {\bibinfo {volume} {76}},\ \bibinfo {pages} {054431} (\bibinfo
  {year} {2007})}\BibitemShut {NoStop}%
\bibitem [{\citenamefont {Stone}\ \emph {et~al.}(2006)\citenamefont {Stone},
  \citenamefont {Zaliznyak}, \citenamefont {Hong}, \citenamefont {Broholm},\
  and\ \citenamefont {Reich}}]{stone2006quasiparticle}%
  \BibitemOpen
  \bibfield  {author} {\bibinfo {author} {\bibfnamefont {M.~B.}\ \bibnamefont
  {Stone}}, \bibinfo {author} {\bibfnamefont {I.~A.}\ \bibnamefont
  {Zaliznyak}}, \bibinfo {author} {\bibfnamefont {T.}~\bibnamefont {Hong}},
  \bibinfo {author} {\bibfnamefont {C.~L.}\ \bibnamefont {Broholm}}, \ and\
  \bibinfo {author} {\bibfnamefont {D.~H.}\ \bibnamefont {Reich}},\ }\href@noop
  {} {\bibfield  {journal} {\bibinfo  {journal} {Nature}\ }\textbf {\bibinfo
  {volume} {440}},\ \bibinfo {pages} {187} (\bibinfo {year}
  {2006})}\BibitemShut {NoStop}%
\bibitem [{\citenamefont {Zhitomirsky}\ and\ \citenamefont
  {Chernyshev}(2013)}]{zhitomirsky2013colloquium}%
  \BibitemOpen
  \bibfield  {author} {\bibinfo {author} {\bibfnamefont {M.}~\bibnamefont
  {Zhitomirsky}}\ and\ \bibinfo {author} {\bibfnamefont {A.}~\bibnamefont
  {Chernyshev}},\ }\href@noop {} {\bibfield  {journal} {\bibinfo  {journal}
  {Rev. Mod. Phys.}\ }\textbf {\bibinfo {volume} {85}},\ \bibinfo {pages} {219}
  (\bibinfo {year} {2013})}\BibitemShut {NoStop}%
\bibitem [{\citenamefont {White}\ \emph {et~al.}(1965)\citenamefont {White},
  \citenamefont {Sparks},\ and\ \citenamefont
  {Ortenburger}}]{white1965diagonalization}%
  \BibitemOpen
  \bibfield  {author} {\bibinfo {author} {\bibfnamefont {R.}~\bibnamefont
  {White}}, \bibinfo {author} {\bibfnamefont {M.}~\bibnamefont {Sparks}}, \
  and\ \bibinfo {author} {\bibfnamefont {I.}~\bibnamefont {Ortenburger}},\
  }\href@noop {} {\bibfield  {journal} {\bibinfo  {journal} {Phys. Rev.}\
  }\textbf {\bibinfo {volume} {139}},\ \bibinfo {pages} {A450} (\bibinfo {year}
  {1965})}\BibitemShut {NoStop}%
\bibitem [{\citenamefont {Oh}\ \emph {et~al.}(2016)\citenamefont {Oh},
  \citenamefont {Le}, \citenamefont {Nahm}, \citenamefont {Sim}, \citenamefont
  {Jeong}, \citenamefont {Perring}, \citenamefont {Woo}, \citenamefont
  {Nakajima}, \citenamefont {Ohira-Kawamura}, \citenamefont {Yamani},
  \citenamefont {Yoshida}, \citenamefont {Eisaki}, \citenamefont {Cheong},
  \citenamefont {Chernyshev},\ and\ \citenamefont {Park}}]{jsoh2016}%
  \BibitemOpen
  \bibfield  {author} {\bibinfo {author} {\bibfnamefont {J.}~\bibnamefont
  {Oh}}, \bibinfo {author} {\bibfnamefont {M.~D.}\ \bibnamefont {Le}}, \bibinfo
  {author} {\bibfnamefont {H.-H.}\ \bibnamefont {Nahm}}, \bibinfo {author}
  {\bibfnamefont {H.}~\bibnamefont {Sim}}, \bibinfo {author} {\bibfnamefont
  {J.}~\bibnamefont {Jeong}}, \bibinfo {author} {\bibfnamefont
  {T.}~\bibnamefont {Perring}}, \bibinfo {author} {\bibfnamefont
  {H.}~\bibnamefont {Woo}}, \bibinfo {author} {\bibfnamefont {K.}~\bibnamefont
  {Nakajima}}, \bibinfo {author} {\bibfnamefont {S.}~\bibnamefont
  {Ohira-Kawamura}}, \bibinfo {author} {\bibfnamefont {Z.}~\bibnamefont
  {Yamani}}, \bibinfo {author} {\bibfnamefont {Y.}~\bibnamefont {Yoshida}},
  \bibinfo {author} {\bibfnamefont {H.}~\bibnamefont {Eisaki}}, \bibinfo
  {author} {\bibfnamefont {S.-W.}\ \bibnamefont {Cheong}}, \bibinfo {author}
  {\bibfnamefont {A.}~\bibnamefont {Chernyshev}}, \ and\ \bibinfo {author}
  {\bibfnamefont {J.-G.}\ \bibnamefont {Park}},\ }\href@noop {} {\bibfield
  {journal} {\bibinfo  {journal} {Nat. Comm. (accepted)}\ } (\bibinfo {year}
  {2016})}\BibitemShut {NoStop}%
\bibitem [{\citenamefont {Aoyama}\ \emph {et~al.}(2013)\citenamefont {Aoyama},
  \citenamefont {Miyake}, \citenamefont {Kagayama}, \citenamefont {Shimizu},\
  and\ \citenamefont {Kimura}}]{aoyama2013pressureEffect}%
  \BibitemOpen
  \bibfield  {author} {\bibinfo {author} {\bibfnamefont {T.}~\bibnamefont
  {Aoyama}}, \bibinfo {author} {\bibfnamefont {A.}~\bibnamefont {Miyake}},
  \bibinfo {author} {\bibfnamefont {T.}~\bibnamefont {Kagayama}}, \bibinfo
  {author} {\bibfnamefont {K.}~\bibnamefont {Shimizu}}, \ and\ \bibinfo
  {author} {\bibfnamefont {T.}~\bibnamefont {Kimura}},\ }\href@noop {}
  {\bibfield  {journal} {\bibinfo  {journal} {Phys. Rev. B}\ }\textbf {\bibinfo
  {volume} {87}},\ \bibinfo {pages} {094401} (\bibinfo {year}
  {2013})}\BibitemShut {NoStop}%
\bibitem [{\citenamefont {Garg}\ \emph {et~al.}(2014)\citenamefont {Garg},
  \citenamefont {Mishra}, \citenamefont {Pandey},\ and\ \citenamefont
  {Sharma}}]{garg2014pressureRaman}%
  \BibitemOpen
  \bibfield  {author} {\bibinfo {author} {\bibfnamefont {A.~B.}\ \bibnamefont
  {Garg}}, \bibinfo {author} {\bibfnamefont {A.}~\bibnamefont {Mishra}},
  \bibinfo {author} {\bibfnamefont {K.}~\bibnamefont {Pandey}}, \ and\ \bibinfo
  {author} {\bibfnamefont {S.~M.}\ \bibnamefont {Sharma}},\ }\href@noop {}
  {\bibfield  {journal} {\bibinfo  {journal} {J. App. Phys.}\ }\textbf
  {\bibinfo {volume} {116}},\ \bibinfo {pages} {133514} (\bibinfo {year}
  {2014})}\BibitemShut {NoStop}%
\bibitem [{\citenamefont {Aronson}\ \emph {et~al.}(1991)\citenamefont
  {Aronson}, \citenamefont {Dierker}, \citenamefont {Dennis}, \citenamefont
  {Cheong},\ and\ \citenamefont {Fisk}}]{aronson1991pressure}%
  \BibitemOpen
  \bibfield  {author} {\bibinfo {author} {\bibfnamefont {M.}~\bibnamefont
  {Aronson}}, \bibinfo {author} {\bibfnamefont {S.}~\bibnamefont {Dierker}},
  \bibinfo {author} {\bibfnamefont {B.}~\bibnamefont {Dennis}}, \bibinfo
  {author} {\bibfnamefont {S.}~\bibnamefont {Cheong}}, \ and\ \bibinfo {author}
  {\bibfnamefont {Z.}~\bibnamefont {Fisk}},\ }\href@noop {} {\bibfield
  {journal} {\bibinfo  {journal} {Phys. Rev. B}\ }\textbf {\bibinfo {volume}
  {44}},\ \bibinfo {pages} {4657} (\bibinfo {year} {1991})}\BibitemShut
  {NoStop}%
\end{thebibliography}%

\end{document}